\documentclass[twocolumn,showpacs,preprintnumbers,prb,superscriptaddress]{revtex4}
\usepackage{graphicx}
\usepackage{amssymb}
\usepackage{amsmath}

\begin{document}

\title{Coulomb drag in graphene single layers separated by thin spacer}
\author{M. I. Katsnelson}
\affiliation{Radboud University of Nijmegen, Institute for
Molecules and Materials, Heyendaalseweg 135, NL-6525 AJ Nijmegen,
The Netherlands}
\date{\today}

\begin{abstract}
Motivated by very recent studies of Coulomb drag in
grahene-BN-graphene system we develop a theory of Coulomb drag for
the Fermi liquid regime, for the case when the ratio of spacer
thickness $d$ to the Fermi wavelength of electrons is arbitrary.
The concentration ($n$) and thickness dependence of the drag
resistivity is changed from $n^{-3}d^{-4}$ for the thick spacer to
$n^{-1}|\ln{(nd^2)}|$ for the thin one.
\end{abstract}


\maketitle

Coulomb drag in bilayer semiconductor systems is a very
interesting phenomenon providing a unique information about
many-body effects \cite{r1,r2,r3,r4}. Since the role of
electron-electron interactions in graphene, a novel
two-dimensional material with extraordinary electronic and
structural properties \cite{rr1,rr2,rr3,rr4,rr5}, is a
controversial issue now (for review, see Ref.\onlinecite{r}) study
of the Coulomb drag in graphene is important, as a way to clarify
the situation.

First theoretical \cite{r5,r6} and experimental \cite{r7,r8}
studies of the Coulomb drag in graphene have been performed
already. Theory \cite{r5} deals with the case of thick spacer
($k_F d \gg 1$, where $k_F$ is the Fermi wave vector of graphene
and $d$ is the spacer thickness), and the results are in a good
agreement with the corresponding experimental data \cite{r7} (the
effects of trigonal warping \cite{r6} seem to be negligible). Very
recently, the group of A. Geim and K. Novoselov has performed
experiments with graphene on a substrate of BN, using also BN as a
spacer between two graphene single layers, and observed the drag
for much thinner spacers \cite{r8}. Here I present a theory
generalizing that of Ref.\onlinecite{r5} for arbitrary values of
$k_F d$.

We start with the general expression for the drag conductivity
\cite{r1,r2} based on the lowest-order perturbation theory in
interlayer Coulomb interaction (we consider only the case of
identical layers 1 and 2, qualitatively all the basic physics
remains the same for the case of different doping of two layers):
\begin{equation}
\sigma _D=\frac 1{16\pi
k_BT}\sum\limits_{\mathbf{q}}\int\limits_{-\infty }^\infty
\frac{d\omega }{\sinh ^2 \left( \frac{\hbar \omega}{2k_BT}\right)
}\Gamma _x^2\left( \mathbf{q},\omega \right) \left| U_{12}\left(
q,\omega \right) \right| ^2  \label{eq1}
\end{equation}
where $\Gamma_x$ is the nonlinear susceptibility of electrons in
the layer, $E_F$ is their Fermi energy, $T$ is the temperature,
$U_{12}$ is the Fourier component of the screened interlayer
Coulomb interaction which reads, for the case of identical layers:
\begin{equation}
U_{12}\left( q,\omega \right) =\frac{u_c\left( q\right) }{\left[
1+v_c\left( q\right) \Pi \left( q,\omega \right) \right] ^2-\left[
u_c\left( q\right) \Pi \left( q,\omega \right) \right] ^2},
\label{eq2}
\end{equation}
$v_c(q)$ and $u_c(q)$ are the Fourier components of bare Coulomb
interactions within the layer and between the layers,
respectively, $\Pi\left( q,\omega \right)$ is the polarization
operator of the electron gas in graphene. In vacuum, $v_c(q)=2\pi
e^2 /q$ and $u_c(q) = v_c(q)\exp{(-qd)}$ and the expression
(\ref{eq2}) coincides with Eq.(A2) from Ref.\onlinecite{r1}.

Since typical frequencies contributing to the integral in
Eq.(\ref{eq1}) is of the order of $k_B T/\hbar$ one can assume,
for low enough temperatures, that the screening is static and
replace $U_{12}\left( q,\omega \right)$ by $U_{12}\left( q,0
\right)$.

In the ballistic regime (which means that the distance between the
layers $d$ is much smaller than the mean-free path within the
layer $l$) one can calculate for the case of graphene \cite{r5}
\begin{equation}
\Gamma _x\left( \mathbf{q},\omega \right) \approx -\frac{4e\tau
\omega }{\pi v}\frac{q_x}q  \label{gamma}
\end{equation}
at $\omega \rightarrow 0$, where $\tau=l/v$ is the mean-free path
time and $v \approx 10^6$ m/s is the Fermi velocity.

Assuming that $\sigma_D \ll \sigma$ where $\sigma$ is the inlayer
conductivity the drag resistivity is $\rho_D = - \sigma_D
/\sigma^2$. Drude formula for the case of graphene is
\begin{equation}
\sigma =\frac{e^2}{\pi \hbar ^2}E_F\tau,   \label{drude}
\end{equation}
thus, the drag resistivity does not depend on $\tau$.

As a result, the drag resistivity for the case of identical
graphene layers takes the form
\begin{equation}
\rho _D=-\frac{2h}{3e^2}\left( \frac{k_BT}{E_F}\right) ^2\sum\limits_{%
\mathbf{q}}\left| \frac{U_{12}\left( q,0\right) }{\hbar v}\right|
^2 \label{eq3}
\end{equation}
The quantity $U_{12}$ in Eq.(\ref{eq3}) is given by the expression
(\ref{eq2}) with the static polarization operator for graphene
(see, e.g., Refs.\onlinecite{r9,r10})
\begin{widetext}
\begin{equation}
\Pi \left( q,0\right) =\left\{
\begin{array}{cc}
\frac{2k_F}{\pi \hbar v}, & q<2k_F \\
\frac{2k_F}{\pi \hbar v}+\frac q{2\pi \hbar v}\left[ \cos ^{-1}\frac{2k_F}q-%
\frac{2k_F}q\sqrt{1-\left( \frac{2k_F}q\right) ^2}\right] , &
q>2k_F
\end{array}
\right.   \label{eq4}
\end{equation}
\end{widetext}

To find $v_c(q)$ and $u_c(q)$ one needs to solve the electrostatic
problem taking into account different screening by substrate,
spacer, and air. Let us assume that the dielectric medium is
three-layer, with the dielectric constant distribution
\begin{equation}
\varepsilon =\left\{
\begin{array}{cc}
\varepsilon _1, & z>d \\
\varepsilon _2, & d>z>0 \\
\varepsilon _3, & z<0
\end{array}
\right.   \label{eps}
\end{equation}
The calculations are quite standard (see, e.g.,
Ref.\onlinecite{polini}). However, for the reader's convenience we
present them here with some details.

We have to solve the Poisson equation
\begin{equation}
\frac d{dz}\left( \varepsilon \left( z\right) \frac{d\varphi
\left( z\right) }{dz}\right) -q^2\varepsilon \left( z\right)
\varphi \left( z\right) =-4\pi e\delta \left( z-\eta \right)
\label{p1}
\end{equation}
where $\varphi \left( z\right) $ is the electrostatic potential
created by the point charge $e$ situated at $x=0,y=0,z=\eta
\rightarrow +0.$ The only allowed solution at $z<0$ is
\begin{equation}
\varphi \left( z\right) =Ae^{qz}  \label{p2}
\end{equation}
and at $z>d$ is
\begin{equation}
\varphi \left( z\right) =Be^{-qz}  \label{p3}
\end{equation}
For $\eta <z<d$ it should be tried in the most general form:
\begin{equation}
\varphi \left( z\right) =\alpha e^{qz}+\beta e^{-qz}  \label{p4}
\end{equation}
From continuity of the potential and the normal component of
electric induction,  $D_n=-\varepsilon \frac{d\varphi }{dz}$ at
the boundaries $z=0$ and $z=d$ we find, taking into account Eqs.
(\ref{p2}) and (\ref{p3}):
\begin{equation}
\frac{\varphi ^{^{\prime }}(-0)}{\varphi (-0)}=\frac{\varepsilon _3}{%
\varepsilon _2}q,  \label{p5}
\end{equation}
\begin{equation}
\frac{\varphi ^{^{\prime }}(d)}{\varphi (d)}=-\frac{\varepsilon _1}{%
\varepsilon _2}q  \label{p6}
\end{equation}
where prime means the derivative with respect to $z$. The
potential $\varphi \left( z\right) $ is continuous at $z=\eta $
($\varphi (-0)=\varphi (+0)$) but its derivative has a jump. Due
to Eqs. (\ref{p1}), (\ref{p2}), (\ref{p5})
\begin{equation}
\varphi ^{^{\prime }}(+0)=\frac{\varepsilon _3}{\varepsilon
_2}qA-\frac{4\pi e}{\varepsilon _2}  \label{p7}
\end{equation}
At last, we can find from Eqs. (\ref{p7}) and (\ref{p6}) the constants $%
\alpha $ and $\beta $. The final answer for $v_c(q)=e\varphi
\left( z=0\right) $ and $u_c(q)=e\varphi \left( z=d\right) $
reads:

\begin{widetext}
\begin{eqnarray}
u_c\left( q\right)  &=&\frac{8\pi e^2\varepsilon _2\exp \left( qd\right) }{%
q\left[ \left( \varepsilon _1+\varepsilon _2\right) \left(
\varepsilon _3+\varepsilon _2\right) \exp \left( 2qd\right)
-\left( \varepsilon _1-\varepsilon _2\right) \left( \varepsilon
_3-\varepsilon _2\right) \right]
},  \nonumber \\
v_c\left( q\right)  &=&\frac{8\pi e^2\varepsilon _2\exp \left(
qd\right) \left[ \varepsilon _2\cosh \left( qd\right) +\varepsilon
_1\sinh \left( qd\right) \right] }{q\left[ \left( \varepsilon
_1+\varepsilon _2\right) \left( \varepsilon _3+\varepsilon
_2\right) \exp \left( 2qd\right) -\left( \varepsilon
_1-\varepsilon _2\right) \left( \varepsilon _3-\varepsilon
_2\right) \right] }.  \label{eq5}
\end{eqnarray}
\end{widetext}
For simplicity, we will consider further only the case
$\varepsilon_1 = \varepsilon_2$ (which, actually, takes place in
the experimental situation \cite{r8} where BN is used both as a
substrate and as a spacer). In this case, the expression
(\ref{eq5}) is simplified dramatically:
\begin{eqnarray} u_c\left( q\right)
&=&v_c\left( q\right) \exp \left( -qd\right) ,  \nonumber
\\
v_c\left( q\right)  &=&\frac{4\pi e^2}{q\left( \varepsilon
_2+\varepsilon _3\right) }  \label{uv}
\end{eqnarray}
and
\begin{equation}
U_{12} =\frac{v_c }{2\left( v_c \Pi  \right) ^2\sinh \left(
qd\right) +\left( 1+2v_c \Pi  \right)\exp \left( qd\right) }.
\label{eq6}
\end{equation}

\begin{figure}[tbp]
\includegraphics[width=9cm]{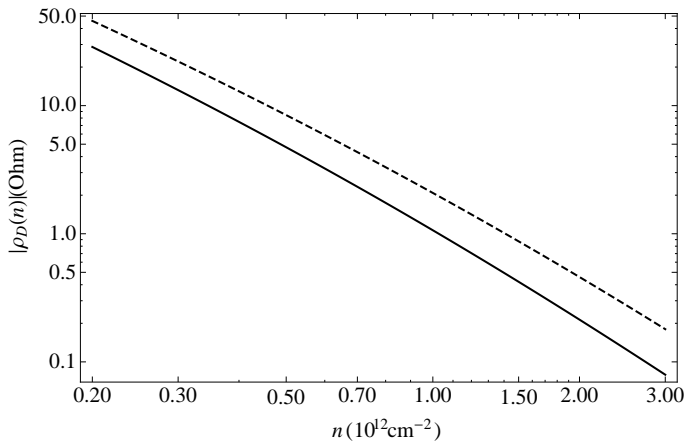}
\caption{Drag resistivity (\ref{eq7}) as a function of charge
carrier concentration, for $\varepsilon_3 =1$, $\varepsilon_2 =
4$, $T$ = 120 K,  $d$ = 3 nm (dashed line) and $d$ = 4 nm (solid
line)} \label{fig:1}
\end{figure}

Substituting Eqs.(\ref{eq4}), (\ref{eq5}), (\ref{eq6}) into
Eq.(\ref{eq3}) we have:
\begin{equation}
\rho _D=-\frac
h{e^2}\frac \pi {48}\left( \frac{k_BT}{E_F}\right) ^2\frac
1{\alpha ^2}F\left( 2k_Fd\right)   \label{eq7}
\end{equation}
where
\begin{equation}
\alpha =\frac{2e^2}{\hbar v\left( \varepsilon _2+\varepsilon
_3\right) } \label{alpha}
\end{equation}
is the effective ``fine structure'' constant (for the combination
of BN with $\varepsilon_2 \approx 4$ and air with $\varepsilon_3
=1$ we have $\alpha \approx 0.87$) and the function $F$ is
represented as
\begin{equation}
F\left( y\right) =\int\limits_0^\infty dx\frac{x^3}{\left[ \varphi
^2\left( x\right) \sinh \left( yx\right) +\frac{x\left( x+4\alpha
\varphi \left( x\right) \right) }{8\alpha ^2}\exp \left( yx\right)
\right] ^2}  \label{eq8}
\end{equation}
where
\begin{equation}
\varphi \left( x\right) =\left\{
\begin{array}{cc}
1, & x<1 \\
1+\frac x2\left( \cos ^{-1}\frac
1x-\frac{\sqrt{x^2-1}}{x^2}\right) , & x>1
\end{array}
\right.   \label{phi}
\end{equation}

In the limit $y \gg 1$
\begin{equation}
F\left( y\right) \cong \frac{3\zeta \left( 3\right) }{2y^4}
\label{eq9}
\end{equation}
and Eqs.(\ref{eq7}), (\ref{eq8}) give the known result \cite{r5}
\begin{equation}
\rho _D=-\frac h{e^2}\frac{\pi \zeta \left( 3\right) }{32}\left( \frac{k_BT}{%
E_F}\right) ^2\frac 1{\left( k_Fd\right) ^2}\frac 1{\left( \kappa
d\right) ^2}  \label{ds}
\end{equation}
where $\kappa = 4 \alpha k_F$ is the inverse screening radius.

In the opposite limit $y \ll 1$ typical values of $x \approx 1/y
\gg 1$ and one can assume $\phi(x) \approx \pi x/4$ which gives
\begin{equation}
F\left( y\right) \cong \left( \frac{8\alpha ^2}{1+\pi \alpha
}\right) ^2\ln \frac 1y  \label{eq10}
\end{equation}

The behavior of drag resistivity as a function of charge carrier
concentration for $k_F d$ of the order of one is shown in Fig. 1.
This result seems to be in a qualitative agreement with the
experimental data \cite{r8}, at least, it gives the correct order
of magnitude for the drag resistivity. At the same time, for small
enough $k_F d$ and $\alpha \approx 1$ the interlayer Coulomb
interaction is in general not small, and it is not clear whether
the lowest-order perturbation theory used here will be also
quantitatively accurate or taking into account next-order
contributions will be necessary. The issue requires further
studies, both experimental and theoretical.

Recently two more works on the subject appeared \cite{ds,cn}. The
results of this paper and the other two papers concerning
concentration dependence of the drag resistivity are in an
agreement, namely, in Ref.\onlinecite{ds} the same analytical
concentration dependence as here, $\rho_D \propto
n^{-1}|\ln{(nd^2)}|$, was obtained for the case of thin spacer
whereas in Ref.\onlinecite{cn} the numerical data were fitted by
$\rho_D \propto n^{-\alpha}$ with $\alpha$ of the order of one.

\section*{Acknowledgement}

I am thankful to Andre Geim for stimulating discussions of
unpublished experimental results \cite{r8} and to Timur
Tudorovskiy for helpful discussions. This work is part of the
research program of the Stichting voor Fundamenteel Onderzoek der
Materie (FOM), which is financially supported by the Nederlandse
Organisatie voor Wetenschappelijk Onderzoek (NWO).


\begin{thebibliography}{99}

\bibitem{r1} A. Kamenev and Y. Oreg, Phys. Rev. B {\bf 52}, 7516 (1995).

\bibitem{r2} K. Flensberg, Ben Yu-Kuang Hu, A.-P. Jauho, and J. M. Kinaret,
Phys. Rev. B {\bf 52}, 14761 (1995).

\bibitem{r3} G. Vignale and A. H. MacDonald, Phys. Rev. Lett. {\bf 76}, 2786 (1996).

\bibitem{r4} J. A. Seamons, C. P. Morath, J. L. Reno, and M. P. Lilly,  Phys. Rev. Lett. {\bf 102}, 026804 (2009).

\bibitem{rr1} A. K. Geim and K. S. Novoselov, Nature Mater. \textbf{6}, 183
(2007).

\bibitem{rr2} M. I. Katsnelson, Mater. Today \textbf{10}, 20 (2007).

\bibitem{rr3} A. H. Castro Neto, F. Guinea, N. M. R. Peres, K. S. Novoselov,
and A. K. Geim, Rev. Mod. Phys. \textbf{81}, 109 (2009).

\bibitem{rr4} A. K. Geim, Science \textbf{324}, 1530 (2009).

\bibitem{rr5} M. A. H. Vozmediano, M. I. Katsnelson, and F. Guinea, Phys.
Rep. {\bf 496}, 109 (2010).

\bibitem{r} V. N. Kotov, B. Uchoa, V. M. Pereira, A. H. Castro
Neto, and F. Guinea, arXiv:1012.3484.

\bibitem{r5} W.-K. Tse, Ben Yu-Kuang Hu, and S. Das Sarma, Phys. Rev. B {\bf 76}, 081401 (2007).

\bibitem{r6} B. N. Narozhny, Phys. Rev. B {\bf 76}, 153409 (2007).

\bibitem{r7} S. Kim, I. Jo, J. Nah, Z. Yao, S. K. Banerjee, and E. Tutuc, Phys. Rev. B {\bf 83}, 161401 (2011).

\bibitem{r8} A. K. Geim, private communication; R. V. Gorbachev, talk at Graphene Week (Obergurgl, April 2011).

\bibitem{r9} T. Ando, J. Phys. Soc. Japan {\bf 75}, 074716 (2006).

\bibitem{r10} B. Wunsch, T. Stauber, F. Sols, and F. Guinea, New J. Phys. {\bf 8}, 318 (2006).

\bibitem{polini} R. E. V. Profumo, M. Polini, R. Asgari, R. Fazio,
and A. H. MacDonald, Phys. Rev. B {\bf 82}, 085443 (2010).

\bibitem{ds} E. H. Hwang and S. Das Sarma, arXiv:1105.3203.

\bibitem{cn} N. M. R. Peres, J. M. B. Lopes dos Santos, and A. H. Castro
Neto, Europhys. Lett. 95, 18001 (2011). 

\end{thebibliography}
\end{document}